# Envelope Control Enabled Probabilistic Shaping for Peak Power Constrained IM-DD Systems

Dongdong Zou, Wei Wang, Jiawen Yao, Zhongxing Tian, Zeyu Feng, Huan Huang, Fan Li, Gordon Ning Liu, Gangxiang Shen, and Yi Cai

*Abstract*—Probabilistic shaping (PS) has attracted significant attention in intensity-modulation and direct-detection (IM-DD) systems. However, due to the unique system model and inherent constraints, the effective application of the PS technique is still an open question in IM-DD systems, particularly in systems with memory effects. In this paper, a novel indirect PS scheme tailored for peak power constrained (PPC) IM-DD systems is proposed. The key idea lies in strategically controlling the signal envelope to mitigate memory-induced impairments, such as nonlinearity, overshoot, peak-to-average power ratio enhancement, etc. The proposed scheme incorporates a dynamic selective mapping (DSLM) mechanism at the transmitter, enabling an untypical bit-to-symbol mapping in which the current symbol is not only determined by the current bits pattern but also by previously generated symbols within a specified memory length. At the receiver side, a turbo equalizer with a modified M-BCJR algorithm is proposed to achieve the recovery of ambiguous bits induced by DSLM. Experimental verification in a 56GBaud PAM8 system demonstrates that the proposed scheme exhibits 1dB receiver sensitivity improvement over 2km single-mode fiber transmission. In addition, the proposed scheme has also been demonstrated to be compatible with the typical probabilistic amplitude shaping architecture, enabling a simple and fine-granularity rate adaptation capability. To the best of our knowledge, this work opens a new sight for the application of the PS technique in PPC IM-DD systems with memory effects.

## I. Introduction

SINCE the inception of optical fiber communication, profound societal changes have transpired within a relatively brief span of several decades. Enabled by optical fiber communication networks, especially data center interconnect (DCI) networks, an increasing number of broadband services are emerging. For short-reach DCI applications, the intensity-modulation and direct-detection (IM-DD) architecture, with its inherent advantages of low cost and power consumption, is deemed a preferable solution over the more expensive coherent technique [1]. For 400Gigabit Ethernet (GbE), 8×50Gb/s or 4×100Gb/s configurations with 4-ary pulse amplitude modulation (PAM4) format are standardized [2]. For 800GbE or 1.6TbE, the discussion also involves PAM4 with a higher single-lane data rate, which implies an increased demand for the bandwidth of optoelectronic devices in transceivers. From another perspective, one can also increase the transmission capacity by enhancing the spectrum efficiency, resulting in extensive research on higher modulation formats, such as PAM6, PAM8, and PAM16 in IM-DD systems [3-5]. With the assistance of the probabilistic shaping (PS) technique, achieving a rate adaptation to match the condition of an actual channel becomes straightforward.

Actually, the concept of PS debuted in 1948 [6] and rekindled in long-haul optical fiber communication systems until 2015 due to the invention of probabilistic amplitude shaping (PAS) structure [7]. In the PAS technique, the constant composition distribution matcher (CCDM) [8] and forward error correction (FEC) encoder are applied in parallel. Thus, both a maximum shaping gain of 1.53dB and fine-granularity rate adaptation can be achieved simply by varying the CCDM rate while keeping the FEC code rate fixed. Due to the superiority of the PS technique, it has been widely investigated in IM-DD optical fiber communication systems. In [9-12], conventional PAS architecture embedded with Maxwell Bolzmann (MB) distribution is directly applied. However, the source of shaping gains needs further exploration as the specific model and constraints of IM-DD systems. In [13], maximum input entropy achieving distribution, specifically exponential distribution, is regarded as the capacity-achieving distribution of an average power constrained (APC) IM-DD system, and a pair-wise MB distribution is proposed to ensure compatibility with the PAS structure [14]. However, the input entropy-maximizing distribution is approximately equal to the capacity-achieving distribution only in a large signal-to-noise ratio (SNR) region. In a more prevalent short-reach IM-DD optical fiber communication system, the absence of optical amplifier implies that the system is peak power constrained (PPC). In such a system, a cup/inverse MB (IvMB) distribution [15, 16] is proposed to acquire the shaping gain based on the intuitive observation that the peripheral symbols suffer fewer disturbances from neighboring symbols compared to the internal symbols. However, this shaping gain experiences a

Manuscript received XXX XXXX; revised XXX, XXXX; accepted XXX, XXXX. This work is partly supported by the National Key R&D Program of China (2022YFB2903000, 2023YFB2906000); National Natural Science Foundation of China (62250710164, 62275185, 62271517, 62035018). (Corresponding Author: Dongdong Zou, and Yi Cai).

Dongdong Zou, Jiawen Yao, Zhongxing Tian, Zeyu Feng, Huan Huang, Gordon Ning Liu, Gangxiang Shen, and Yi Cai are with the School of Electronic and Information Engineering, Soochow University, Suzhou 215006, China. (e-mail: ddzou@suda.edu.cn, yicai@suda.edu.cn)

Wei Wang, Fan Li are with the School of Electronics and Information Technology, Guangdong Provincial Key Laboratory of Optoelectronic Information Processing Chips and Systems, Sun Yat-Sen University, Guangzhou 510275, China (e-mail: lifan39@mail.sysu.edu.cn).

decline with the increase in modulation format, attributed to the reduced proportion of peripheral symbols. Besides, this distribution appears to be more vulnerable to practical system impairments, such as modulation nonlinearity. In [17], a comprehensive discussion of the shaping gain in PPC IM-DD systems is provided. It is emphasized that the peak-to-average power ratio (PAPR) enhancement is a pivotal parameter for the PS design in this system. In our previous work [18], the capacity-achieving distribution of memoryless IM-DD systems is numerically solved by the Blahut-Arimoto (BA) algorithm. According to the results, the shaping gain in memoryless PPC IM-DD systems appears only at the low achieved information rate (AIR) regions, rendering it ineffective in future high spectral efficiency systems.

This paper is an extended version of our previous work [19] published in OFC 2025. A comprehensive discussion on the implementation of the PS technique in PPC IM-DD systems is presented. First of all, the achievable shaping gain comparison between several commonly applied distributions, including MB, inverse MB, and BA-solved capacity-achieving distributions, is provided in the memoryless PPC IM-DD systems. According to the results, there is no shaping gain for the MB distribution, and only a slight shaping gain for the inverse MB and BA-solved distributions in the low AIR regions. Moreover, from the perspective of envelope control, a novel indirect PS scheme is tailored for the PPC IM-DD systems exhibiting memory effects. In this scheme, a dynamic selective mapping (DSLM) algorithm is devised to avoid the presence of specific 'bad' symbol patterns at the transmitter side, consequently alleviating certain impairments introduced by system memory, such as nonlinearity, overshoot, and PAPR enhancement. At the receiver, a modified turbo equalizer is proposed to recover the ambiguous bits induced by DSLM. The effectiveness of the proposed scheme is experimentally validated in a 56GBaud PAM8 system, and the results indicate that about 1dB receiver sensitivity enhancement is achieved over a 2km single-mode fiber (SMF) transmission. Finally, the proposed scheme is also demonstrated to be compatible with the PAS structure, enabling envelope control of any shaped signal and exhibiting a simple and fine-granularity rate adaptation capability.

The remainder of this paper is organized as follows. Section II elaborates on the foundational system model of PPC IM-DD systems and gives a brief introduction to the BA algorithm. Furthermore, the underlying principle of the proposed DSLM and modified turbo equalizer is also presented in this part. Afterward, the numerical simulation, experimental investigation, and results discussion will be presented in Section III. Finally, the paper will be concluded in Section IV.

## II. SYSTEM MODEL AND METHODS

Information theory serves as the bedrock of communication science and remains an enduring topic in this field. Channel capacity, as one of the major focuses, is extensively studied in diverse systems. According to the definition of channel capacity, it can be expressed as:

$$C = \max_{p(x)} I(X;Y). \quad (1)$$

It means the maximum achievable mutual information between the transmitted signal $X$ and the received signal $Y$. Where $p(x)$ is the probability mass function (PMF) of the input signal $X$. For a linear long-haul coherent communication system with optical amplifiers to boost the optical signal into a desired launch power or compensate for the transmission loss, it is always regarded as the well-known memoryless additive white Gaussian noise (AWGN) model with the average optical power constraint. In this model, the channel can be well characterized by:

$$p(y \mid x) = \frac{1}{\sqrt{2\pi\sigma^2}} \exp(-\frac{\|y-x\|^2}{2\sigma^2}). \quad (2)$$

$$E\left[|l \cdot X|^2\right] = I_{ave}. \quad (3)$$

where $p(y \mid x)$ is the conditional probability density function, $\sigma^2$ is the variance of the Gaussian noise, $E[\cdot]$ is the mean operation, $l$ is a signal scaling factor, and $I_{ave}$ is the limited average optical power. For this system, the capacity-achieving distribution can be easily solved by the Lagrange multiplier method, and the solution is the Gaussian distribution [6]. For a finite support constellation size $M$, the distribution is the well-known MB distribution. However, owing to the fundamental differences in system models and constraints associated with IM-DD systems compared to coherent communication systems, MB distribution cannot be directly applied in IM-DD systems.

In this section, a detailed system model for the PPC IM-DD system is established. Subsequently, the classical BA algorithm is briefly introduced to iteratively solve the capacity-achieving distribution for a linear memoryless PPC IM-DD system. Finally, a novel indirect PS technique is tailored for a practical PPC IM-DD system with memory effects.

### A. System Model of PPC IM-DD Systems

IM-DD architecture has always been regarded as the preferred solution for short-reach applications, particularly for systems with transmission distances below 2km, such as optical intra-DCI. In such an application, the optical amplifier is not equipped due to the negligible transmission loss and system cost constraints. Hence, the PPC is valid, and the system performance is predominantly determined by the receiver side white Gaussian noise, including the thermal noise of photodiode (PD) and radio frequency amplifiers such as the transimpedance amplifier (TIA). The detailed channel model for PPC IM-DD systems has been given in Fig. 1. The nonnegative transmitted sequence $X$ is nonlinearly modulated into the optical domain, namely, linear modulation from the

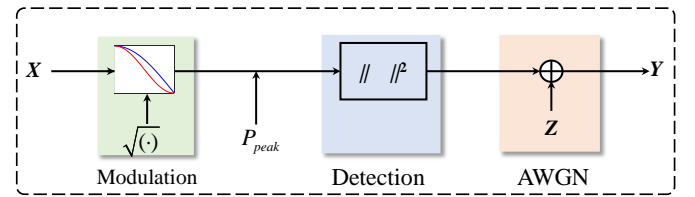

Fig. 1. Channel model of peak power constrained IM-DD systems.

electric field to optical intensity. We denote the optical field signal by:

$$X' = \sqrt{X}. \quad (4)$$

Due to the requirement of modulation linearity during the electric-to-optical conversion, the amplitude of the electrical signal, namely the peak power of the optical signal, is limited, which can be expressed as:

$$\max[X']^2 = \max[X] \leq P_{peak}. \quad (5)$$

After the fiber transmission, the signal is detected by a single-end PD with square-law detection. As the system is dominated by the receiver side noise, such as thermal noise from PD and TIA, an AWGN is added after the PD. Thus, the system model can be established as:

$$Y = X + Z. \quad (6)$$

$$s.t. \begin{cases} \max[X] < P_{peak} \\ X \geq 0 \end{cases}. \quad (7)$$

For a more specific expression, the mutual information can be written as:

$$I(X;Y) = H(Y) - H(Y|X). \quad (8)$$

Due to the independent characteristics between transmitted signal $X$ and noise $Z$, the system capacity can be rewritten as:

$$C = \max_{p(x)} I(X;Y) = \max_{p(x)} \{H(Y)\} - \frac{1}{2}\log_2(2\pi e\sigma^2). \quad (9)$$

And the system is constrained by Eq. (7).

*B. Blahut-Arimoto Algorithm*

In contrast to long-haul coherent systems, closed-form analytical solutions are often intractable in many communication systems, such as PPC and APC IM-DD systems. In 1972, a numerical programming method, namely the BA algorithm [20, 21], was proposed to solve the system capacity and capacity-achieving distribution in an iterative way. In this algorithm, the capacity $C$ is denoted as:

$$C = \max_{p(x)} \sum_X \sum_Y p(x)p(y|x)\log_2 \frac{p(x|y)}{p(x)}. \quad (10)$$

where $p(x|y)$ is the posterior probability function, and it is determined by the priori probability $p(x)$ and conditional probability $p(y|x)$:

$$p(x|y) = \frac{p(x)p(y|x)}{\sum_X p(x)p(y|x)}. \quad (11)$$

Utilizing the Lagrange multiplier method, the cost function can be expressed as:

$$J = \sum_X \sum_Y p(x)p(y|x)\log_2 \frac{p(x|y)}{p(x)} - \lambda\left[\sum_X p(x) - 1\right]. \quad (12)$$

If any other system constraint is considered, it can be added in Eq. (12). Taking the partial derivative of the cost function with respect to $p(x)$,

$$\frac{\partial J}{\partial p(x)} = \sum_Y \{p(y|x) \cdot \log_2 p(x|y)\} - \log_2 p(x) - 1 + \lambda = 0. \quad (13)$$

The solution of $p(x)$ can be obtained as:

$$p(x) = \frac{2^{\sum_Y p(y|x) \cdot \log_2[p(x|y)]}}{\sum_X \left\{2^{\sum_Y p(y|x) \cdot \log_2[p(x|y)]}\right\}}. \quad (14)$$

The solution of $p(x)$ is determined by conditional probability $p(y|x)$ and posterior probability $p(x|y)$. Actually, the solving process of the BA algorithm is an alternating optimization procedure. Given $p(y|x)$ and $p(x)$, the $p(x|y)$ that maximizes the capacity can be calculated by Eq. (11). Subsequently, for the given $p(y|x)$ and $p(x|y)$, the $p(x)$ that maximizes the capacity can be determined by Eq. (14). It has been proven that when the iterative number goes to infinity, the $p(x)$ will approach the capacity-achieving distribution, and the mutual information approaches the channel capacity [20, 21].

*C. Dynamic Selective Mapping Algorithm*

According to the discussion in [17], it seems that a sensible solution for applying the PS technique in the PPC IM-DD system should take the system memory effects into consideration. In this paper, we propose a dynamic bit-to-symbol mapping scheme to achieve envelope control of the transmitted signal, consequently eliminating system memory-induced impairments, such as nonlinearity, overshoot, and PAPR enhancement. The detailed scheme diagram of the DSLM strategy is shown in Fig. 2. In a typical PAM system, the original bit sequence $B_0$ is processed by the FEC encoder to add the redundant bits. After interleaving, the bit-to-symbol mapping is realized by the Gray code with the one-to-one mapping rule. In the proposed DSLM scheme, the final mapped symbol $\hat{s}_k$ is not only determined by the corresponding bit pattern $[b_{km-m+1},...,b_{km}]$ according to the

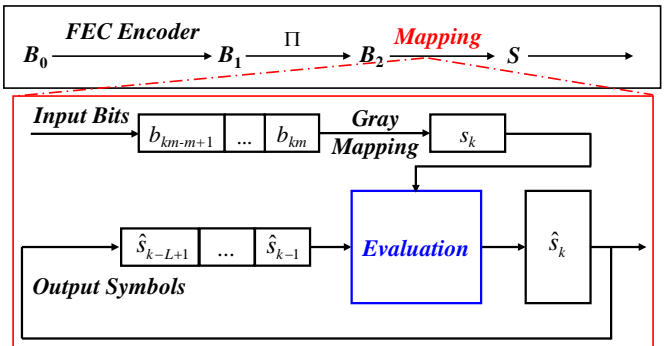

Fig. 2. Scheme diagram of dynamic selective mapping.

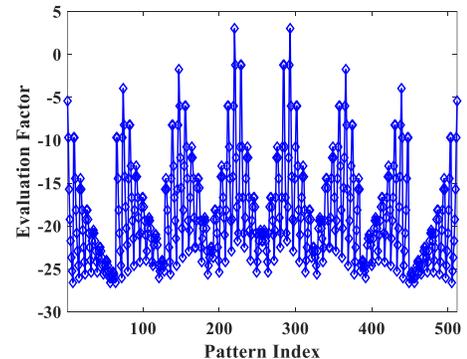

Fig. 3. Evaluation factor of different PAM8 symbol sequence pattern ($L$=5).

Gray mapping but also impacted by the previously mapped symbol sequence $[\hat{s}_{k-L+1},...,\hat{s}_{k-1}]$ within a desired memory length $L$.

Since the purpose of the proposed algorithm is to control the envelope of the transmitted signal, one of the critical issues is how to evaluate the quality of a symbol sequence pattern, namely the signal envelope within a short memory length. Typically speaking, a symbol sequence can be evaluated by its statistical properties or with the assistance of the channel state information (CSI). In this paper, we assume that the CSI is unavailable, and the symbol sequence is evaluated as:

$$f(S) = -10\cdot \log_{10}\{var(S)\cdot \max[abs(S)]\}. \quad (15)$$

where $S$ is the symbol sequence pattern with the length of $L$. $var(\cdot)$ denotes the variance operation. The evaluation function is inspired by the overshoot phenomenon, as the symbol sequences with large peak amplitude and significant fluctuations are more prone to the overshoot issue. According to Eq. (15), the evaluation factor of different PAM8 symbol sequence patterns ($L$=5) is depicted in Fig. 3. Moreover, in the scenarios where the partial CSI is available at the transmitter side, the sequence pattern can be evaluated with the assistance of CSI, such as:

$$f(S) = S * h \quad (16)$$

Once a reliable evaluation function has been identified, another key step is to determine how to transform a 'bad' symbol sequence pattern into a 'good' desired one. Algorithm 1 gives out the utilized strategy in this paper:

---

**Algorithm 1.** Dynamic Selective Mapping

**Initiation:** Import the modulation format $M$, considering memory length $L$, and sequence forbidden ratio $\gamma$, and dividing all possible symbol sequence patterns into $F$ and $\bar{F}$.
**for** $k$ in 1 to $N$ **do**         // $N$ is the length of the output symbol
 $s_k = Gray\{[b_{km-m+1},...,b_{km}],M\}$;    // Gray mapping
 **if** $[\hat{s}_{k-L+1},...,\hat{s}_{k-1},s_k] \in \bar{F}$
  $\hat{s}_k = s_k$;
 **else**
  find $A \subseteq \chi \to \forall a \in A, [\hat{s}_{k-L+1},...,\hat{s}_{k-1},a] \in \bar{F}$;
  **if** $A = \varnothing$
   $\hat{s}_k = s_k$;
  **else**
   find $C \subseteq A \to \forall c \in C, c = \arg\min_{x\in A}[d_H(x,s_k)]$;
   **if** $|C| = 1$         // $|C|$ returns the element number of $C$
    $\hat{s}_k = c$;
   **else**
    find $D \subseteq C \to \forall d \in D, d = \arg\min_{x\in C}[f([\hat{s}_{k-L+1},...,\hat{s}_{k-1},x])]$;
    **if** $|D| = 1$
     $\hat{s}_k = d$;
    **else**
     find $G \subseteq D \to \forall g \in G, g = \arg\min_{x\in D}[d_E(x,s_k)]$;
     $r = RandomInteger(1,|G|)$;    // Random Index
     $\hat{s}_k = G(r)$;
    **end if**
   **end if**
  **end if**
 **end if**
**end for**

---

**Initialization:** Input the modulation format $M$, symbol set $\chi = [-M+1,...,M-1]$, considering memory length $L$, and a pattern forbidden ratio $\gamma$.

1. Mapping the current bit pattern $[b_{km-m+1},...,b_{km}]$ into the symbol $s_k$ according to the Gray code. Dividing all the sequence patterns into two groups, namely the forbidden group $F$ and the non-forbidden group $\bar{F}$ according to considering memory length $L$, evaluation function Eq. (15), and forbidden ratio $\gamma$.
2. If the current symbol sequence pattern $S^k_{k-L+1} = [\hat{s}_{k-L+1},...,\hat{s}_{k-1},s_k]$ does not belong to group $F$, the current symbols $s_k$ can be directly transmitted. Otherwise, go to step 3.
3. Finding symbol $a \in \chi$ such that the sequence pattern $[S^{k-1}_{k-L+1},a] = [\hat{s}_{k-L+1},...,\hat{s}_{k-1},a]$ belongs to the group $\bar{F}$. All symbols that satisfy this specified criterion constitute subset $A$. If $A = \varnothing$, current symbol $s_k$ can be directly transmitted. Otherwise, go to step 4.
4. Identifying symbol $c$ ($c \in A$) exhibits the minimum Hamming distance to $s_k$. All symbols that meet the criteria constitute subset $C$. If only a single element $c$ is contained in $C$, the $c$ is regarded as the output of this stage, namely $\hat{s}_k = c$. Otherwise, go to step 5.
5. Calculating the evaluation factor of sequence pattern $[\hat{s}_{k-L+1},...,\hat{s}_k,x]$, in which $x \in C$, and selecting the symbol provides the best evaluation factor. All symbols that satisfy this criteria form subset $D$. If only a single element $d$ is contained in $D$, the $d$ is regarded as the output of this stage, namely $\hat{s}_k = d$. Otherwise, go to step 6.

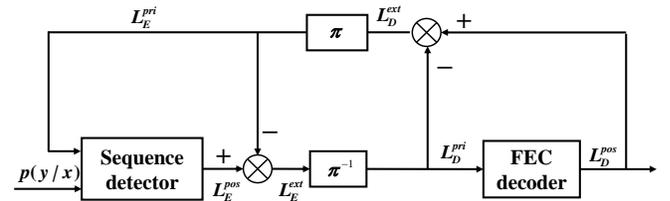

Fig. 4. Scheme diagram of turbo equalization.

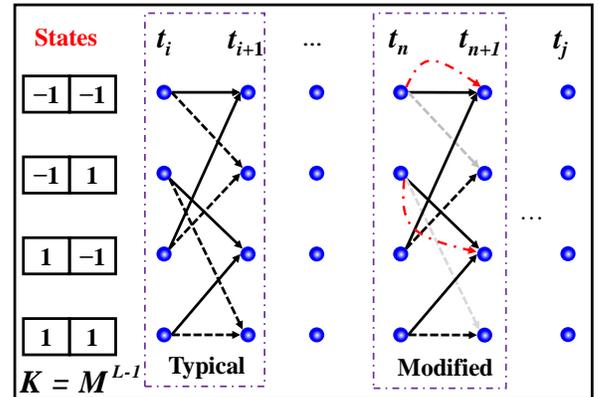

Fig. 5. Typical trellis and modified trellis in the sequence detector. In this figure, the modulation format is OOK, and the memory length is 3. The modified state transitions are only used for illustration, and not the actual state transitions.

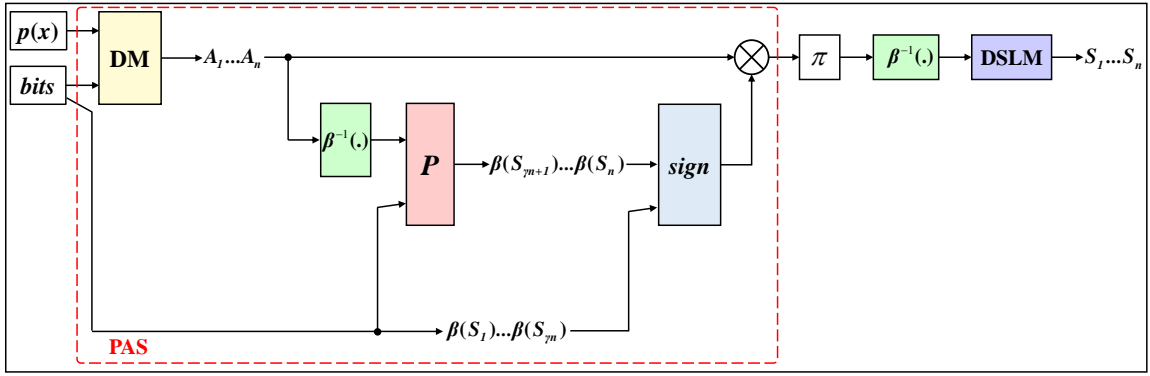

Fig. 6. Scheme diagram of PAS coupled by DSLM to realize rate adaptation and envelope control of shaped signal.

6. Determining the symbol within $D$ that exhibits the minimum Euclidean distance to $s_k$. All symbols that satisfy this requirement constitute subset $G$. The final transmitted symbol can be randomly selected from $G$.

*D. Turbo Equalization and Modified M-BCJR Detector*

As described in subsection *C* of section II, the essence of the proposed envelope control scheme lies in a dynamic bit-to-symbol mapping rather than a traditional one-to-one mapping. Thus, a straightforward symbol-to-bit de-mapping at the receiver will induce a significant bit ambiguity. In order to deal with this issue, the turbo equalization scheme with a modified M-BCJR algorithm is proposed. Fig. 4 gives the diagram of the turbo equalization scheme, in which the main idea is to utilize the soft information between the two soft input and soft output modules, namely the sequence detector and the FEC decoder, to enhance the performance of the detector and decoder iteratively. In this paper, the DVB-S2 low-density parity check (LDPC) provided by MATLAB is directly utilized as the system FEC codes. For the soft output sequence detector, the maximum a posteriori probability (MAP) detector based on the BCJR algorithm is employed, but a necessary modification tailored for the DSLM is performed. Fig. 5 shows the trellis representation of a Markov chain with modulation format $M = 2$ and memory length $L = 3$. At stage $i-1$, the current state is expressed by the symbol pattern with length of $L-1$, i.e. $[s_{i-L+1},...,s_{i-1}]$. Hence, the number of all states is $K = M^{L-1}$. In a traditional trellis, each state at stage $i$ is derived from $M$ distinct states at stage $i-1$ and will transition to $M$ distinct states at stage $i+1$ as shown in Fig. 5. In the proposed scheme, the state transition relationship is modified due to the 'bad' pattern suppression mechanism, which is also expressed in Fig. 5 at stage $n$. For a considering memory length $L$ and forbidden ratio $\gamma$, the trellis with a specific state transition relationship can be regarded as a priori information at the receiver side. Within this state transition framework, the forward recursion and backward recursion can be successfully achieved. In order to reduce the computation complexity of the BCJR detector, the M-algorithm [22, 23] is adopted in this paper.

*E. Rate Adaptability Consideration*

In long-haul coherent communication systems, PS technique has achieved remarkable success in recent years, attributed to not only the maximum 1.53dB shaping gain but also the low complexity and fine-granularity rate adaptive capability introduced by the PAS framework. In this paper, we aim to achieve the envelope control of the transmitted signal by manipulating the bit-to-symbol mapping, which can be regarded as an indirect probabilistic shaping approach, as the final mapped signal shows a non-uniform distribution. If the input bits of the DSLM module originate from the information source directly, it is challenging to realize a fine-granularity rate adaptation. In this paper, we propose a cascade structure, including PAS and DSLM, as shown in Fig. 6, to realize the rate adaptation and envelope control. The bit source of the DSLM module is de-mapped from the output symbol of the PAS unit, followed by an interleaver. As the PAS block is only utilized for low complexity rate adaptation, any PAS-compatible distribution, such as MB, IvMB, twin exponential, BA-solved capacity-achieving distribution, etc, can be employed with specific consideration.

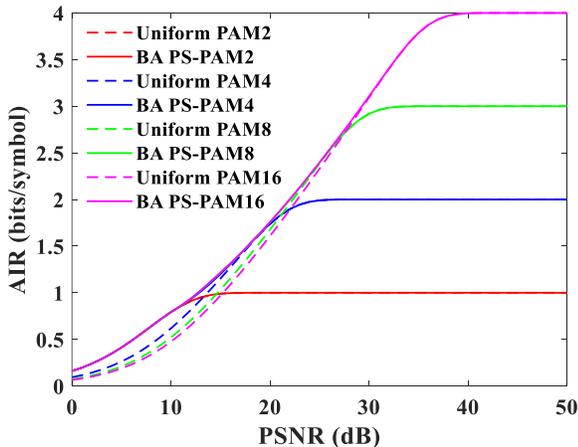

Fig. 7. AIR versus PSNR in memoryless PPC IM-DD systems.

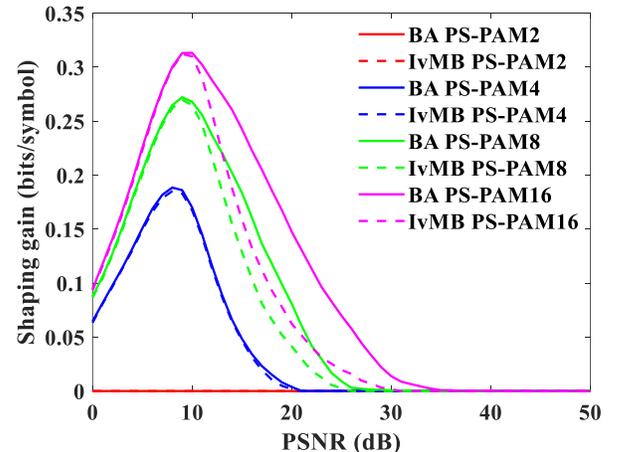

Fig. 8. Shaping gain versus PSNR in memoryless PPC IM-DD systems.

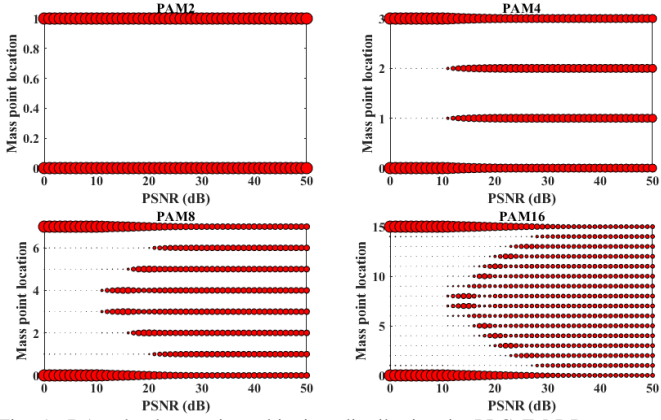

Fig. 9. BA-solved capacity-achieving distribution in PPC IM-DD systems. Size of the particle indicates probability.

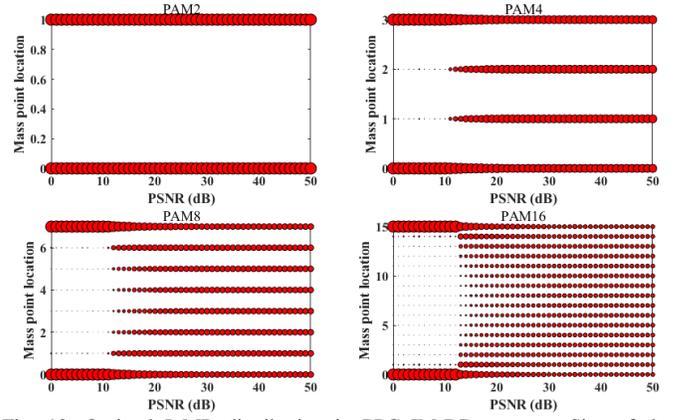

Fig. 10. Optimal IvMB distribution in PPC IM-DD systems. Size of the particle indicates probability.

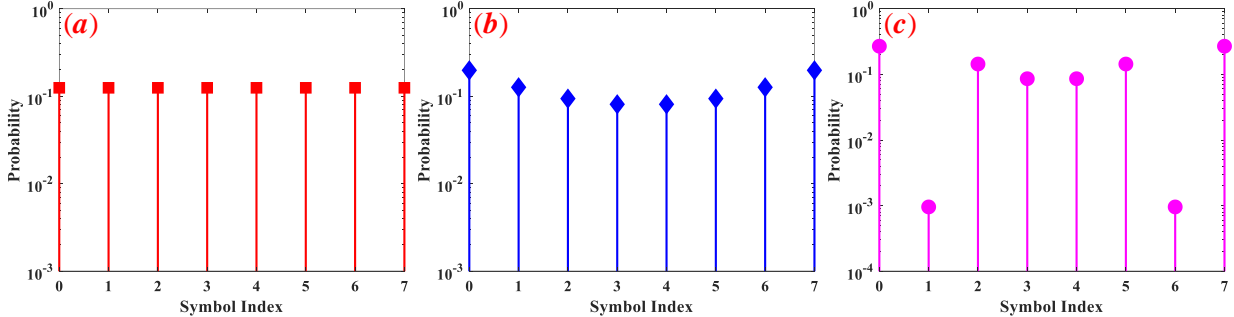

Fig. 11. Optimal distribution for (a) MB PS-PAM8, (b) IvMB PS-PAM8, and (c) BA-solved PS-PAM8 at 20dB PSNR.

## III. RESULTS AND DISCUSSION

In this section, we first numerically evaluate the shaping gain of several distributions in the memoryless PPC IM-DD system, including MB, IvMB, and BA-solved capacity-achieving distributions. Then, the experimental demonstration of the proposed novel PS technique, i.e. DSLM scheme, is provided in a PPC PAM8 system.

### A. Results for Linear Memoryless PPC IM-DD Systems

For PPC IM-DD systems, several distributions have been discussed to obtain the shaping gain by experiments. However, it is rather rare to discuss the shaping gain of these distributions from the perspective of information theory. In this manuscript, we first investigate the shaping gain of MB, IvMB, and BA-solved capacity-achieving distributions in a memoryless PPC IM-DD system. A detailed application of the BA algorithm for capacity solving can refer to our previous publication [18]. Fig. 7 gives the AIR versus the peak-signal-to-noise ratio (PSNR) of several specific modulation formats, including OOK, PAM4, PAM8, and PAM16. According to Fig. 7, there is no shaping gain for the OOK modulation format. For higher modulation formats, the shaping gain is evident in the low PSNR/AIR region, which falls outside the scope of future high spectral efficiency communication systems. For MB and IvMB distributions, the shaping gain is obtained by the Monte Carlo simulation. According to the simulation results, there is no shaping gain for the MB distribution over all of the modulation formats, as the optimal input entropy at any PSNR is always $\log_2 M$. The detailed shaping gains for BA-solved capacity-achieving distribution and IvMB distribution are given in Fig. 8. It can be found that the shaping gain of BA PS-PAM4 and IvMB PS-PAM4 is almost the same. For PAM8 and PAM16 modulation formats, the BA-solved distribution and IvMB distribution exhibit similar shaping gain at the low PSNR regions, such as PSNR<10dB, while a significant shaping gain gap between the two distributions is observed at the moderate PSNR regions. It can also be seen from Fig. 8 that the peak shaping gain for PAM4, PAM8, and PAM16 modulation formats is observed with the PSNR slightly lower than 10dB. Under these PSNR regions, the system AIR is generally lower than 1 bit/symbol, resulting in the shaping gain being invalid, since these regions will not be involved in the high-speed communication systems. Furthermore, for the PAM8 modulation format with 2 bits/symbol AIR, corresponding to around 22dB PSNR, the shaping gain is only 0.0566 bits/symbol. For the PAM16 modulation format with 3 bits/symbol AIR, the available shaping gain is only 0.0237 bits/symbol. Figs. 9 and 10 give out the capacity-achieving distribution and optimal IvMB distribution versus the PSNR for several modulation formats, in which the size of the particle indicates probability. Both the BA-solved capacity-achieving distribution and the optimal IvMB distribution exhibit a strong concentration of probability on the two symmetric outermost symbols under low PSNR conditions. For high PSNR regions, the BA-solved capacity-achieving distribution and optimal IvMB distribution turn to uniform distribution as the PSNR is large enough to achieve the finite AIR of $\log_2 M$. A significant difference between the two distributions is observed in the moderate PSNR regions. Specifically, the BA-solved capacity-achieving distribution shows a non-monotonic symmetric distribution. For a more intuitive exhibition, the PDFs of optimal MB PS-PAM8, IvMB PS-PAM8, and BA-solved PS-PAM8 at 20dB PSNR are depicted in Fig. 11.

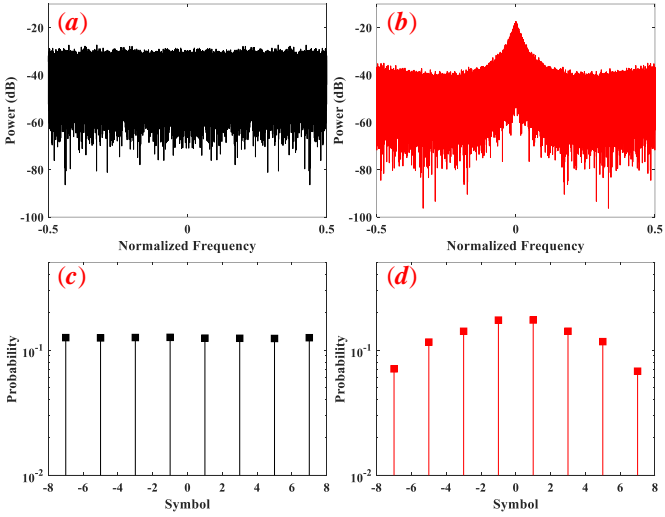

Fig. 12. Electrical spectrum of (a) uniform PAM8, (b) DSLM PAM8. Distributions of (c) uniform PAM8, (d) DSLM PAM8. For DSLM PAM8, the memory length $L$ and forbidden ratio $\gamma$ are 5 and 0.8, respectively.

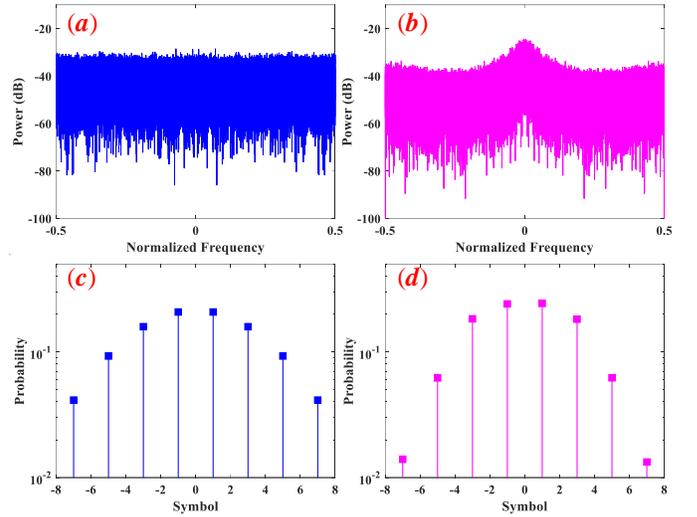

Fig. 13. Electrical spectrum of (a) MB-PS PAM8 (H=2.8), (b) DSLM MB-PS PAM8. Distributions of (c) MB-PS PAM8, (d) DSLM MB-PS PAM8. For DSLM MB-PS PAM8, the memory length $L$ and forbidden ratio $\gamma$ are 5 and 0.8, respectively.

*B. Results for Practical PPC IM-DD Systems*

According to the results presented in subsection A of section III, the theoretical shaping gain in a high spectral efficiency memoryless PPC IM-DD system is negligible. In this paper, we diverge from the traditional concept of PS, proposing a novel indirect PS technique, in which the envelope of the transmitted signal is controlled to eliminate the system memory-induced impairments. To validate the effectiveness of the proposed scheme, bits demapped from PAM8 symbols with different distributions are injected into the DSLM module. Figs. 12(a) and 12(b) illustrate the spectrum of the DSLM output signal with memory length $L=5$, where the input bits are demapped from uniform PAM8, under forbidden ratios of 0 and 0.8, respectively. The corresponding distributions of the DSLM output symbols are exhibited in Figs. 12(c) and 12(d). For the case with uniform input and forbidden ratio of 0, the DSLM is regarded as an all-pass mechanism, resulting in a flat spectrum and uniform output as shown in Figs. 12(a) and 12(c). For the case of 0.8 forbidden ratio, the output symbols do not follow an independent and identically distribution (i.i.d.), and a low-pass filtering phenomenon is easily observed in Fig. 12(b). The corresponding symbol distribution is presented in Fig. 12(d), which shows a cap shape. This phenomenon is predictable as the peak power is a vital factor in the predefined evaluation function, resulting in a higher suppression probability of the symbol with large power. Based on this observation, a cap-shaped MB distribution (designated as MB distribution) rather than a cup-shaped MB distribution (designated as IvMB distribution) is employed in the PAS framework in following rate adaptative mechanism discussion. Fig. 13(a) shows the spectrum of the DSML output signal when the input bits are demapped from the MB-PS PAM8 symbol (H=2.8) and the forbidden ratio is 0. As the forbidden ratio is 0, the output symbol of DSLM is the same as the output of PAS, namely the standard MB-PS PAM8. When the forbidden ratio increases to 0.8, the spectrum and distribution of the DSLM output symbol are presented in Figs. 13(b) and 13(d). It is evident that a low-pass filtering effect is observed, accompanied by a reduction in the probability of symbols with larger amplitudes.

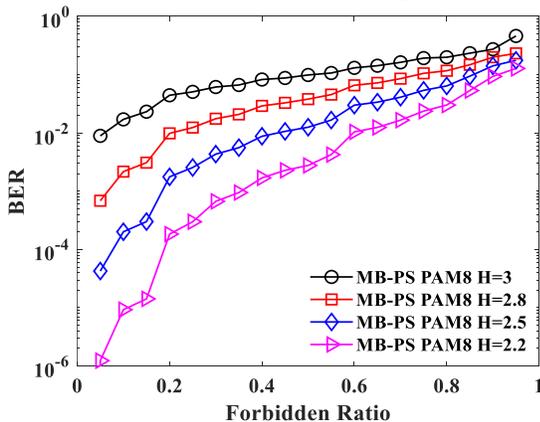

Fig. 14. DSLM induced BER (direct decision) versus forbidden ratio with different input.

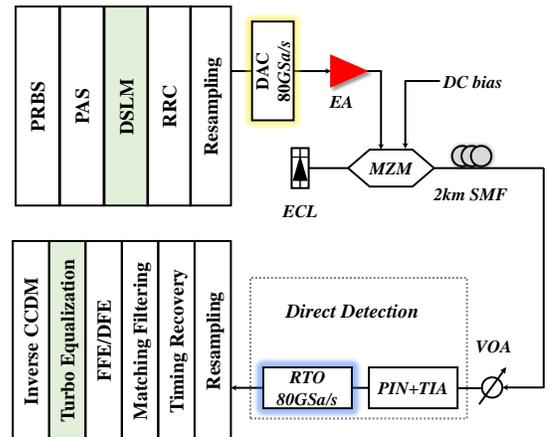

Fig. 15. DSP and experimental setup.

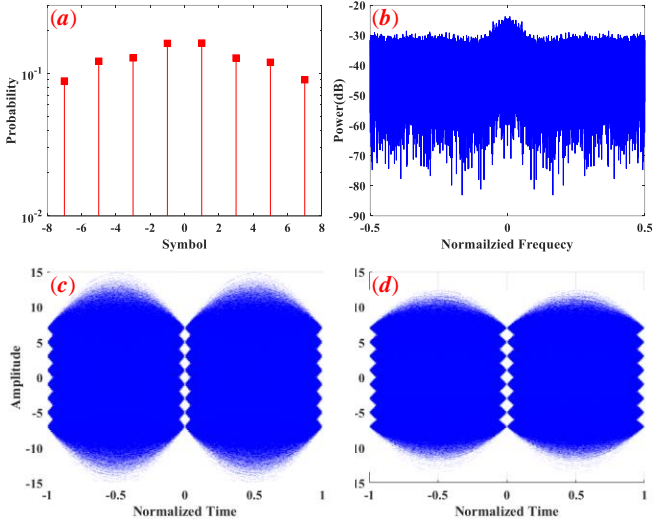

Fig. 16. (a) Symbol distribution and (b) spectrum of DSLM PAM8 ($L=5$, forbidden ratio $\gamma = 0.33$), (c) eye diagram of uniform PAM8 and (d) DSLM PAM8 with 100 up-sampling rate.

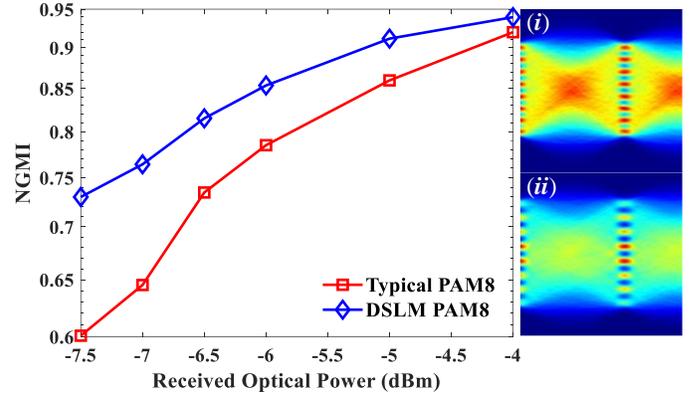

Fig. 17. NGMI versus ROP of 56GBaud uniform PAM8 and DSLM PAM8. Eye diagrams of equalized (i) uniform PAM8, (ii) DSLM PAM8 at −4dBm ROP.

It is anticipated that the shaping gain of the DSLM scheme increases with the forbidden ratio. However, this shaping gain is achieved at the cost of introducing ambiguous bits, which directly leads to bit errors. Fig. 14 shows the direct decision bit error rate (BER) versus the forbidden ratio of MB-PS PAM8 with different entropies. It is evident that an increase in the forbidden ratio leads to a corresponding rise in the BER induced by DSLM. Moreover, under identical forbidden ratios, the BER caused by DSLM is lower in MB-PS PAM8 with reduced entropy compared to those with higher entropy. In this paper, the DSLM induced bit ambiguity issue is handled by the turbo equalization technique. For a practical system with a determined FEC scheme, the optimal forbidden ratio can be determined by adjusting the BER, induced by DSLM, to approach the FEC threshold.

To verify the effectiveness of the proposed scheme, an experimental demonstration is carried out in a PAM8 system. Fig. 15 shows the detailed experimental setup and digital signal processing (DSP) at transceivers. A pseudo-random bit sequence (PRBS) is generated as the bit source. After that, the PAS unit is applied to transform the bit sequence to the PAM8 symbol with a desired distribution. Within the PAS module, the DVB-S2 LDPC with a 2/3 code rate is applied as the FEC codes. Before the DSLM procedure, an interleaver and symbol-to-bits de-mapping based on Gray rules are required. Then, the signal is shaped by a raised cosine filter (RRC) and resampled to match the sampling rate of the digital-to-analog converter (DAC).

The offline generated signal is uploaded into a Fujitsu DAC with 8-bit resolution and 80GSa/s sampling rate, and boosted by an SHF electric amplifier (EA) with 30GHz bandwidth and 24dB gain. After that, the signal's electrical-to-optical (EO) conversion is realized in an intensity Mach-Zehnder modulator (MZM-7939), in which the optical carrier with 1550nm wavelength and 16dBm power is generated from an external cavity laser (ECL). The optical signal is launched into a 2km single mode fiber (SMF) with 6dBm optical power. At the receiver, the received optical power is controlled by a variable optical attenuator (VOA), and the signal is detected by a 40Gbps PIN PD. Finally, the signal is captured by a real-time oscilloscope (RTO) with 36GHz bandwidth and 80GSa/s sampling rate.

In the receiver side offline DSP, the signal is resampled to 2sps, and the timing phase recovery is performed. After that, matched filtering is applied to achieve the maximum system SNR. A typical feedforward equalizer (FFE) or decision feedback equalizer (DFE) is employed to compensate for the imperfect channel response. Subsequently, turbo equalization is utilized to eliminate the DSLM induced ambiguous bits. Finally, the decoded bit should be mapped into the PAM8 symbol based on the Gray code, and an inverse CCDM is required to transform the PAM8 symbol to the original bit sequence.

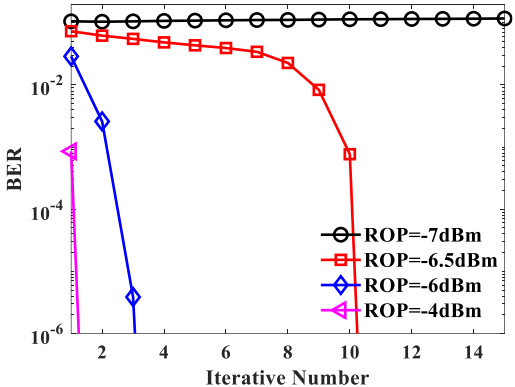

Fig. 18. BER versus iterative number of turbo equalization in DSLM PAM8 system.

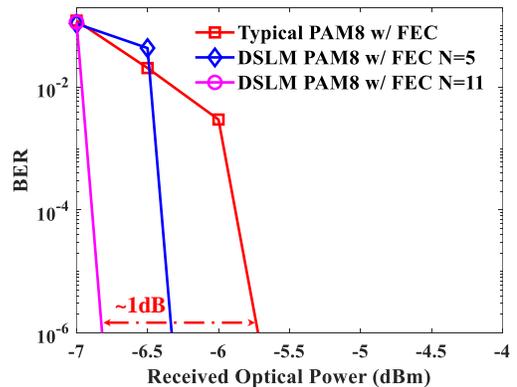

Fig. 19. BER versus received optical power in typical uniform PAM8 and DSLM PAM8.

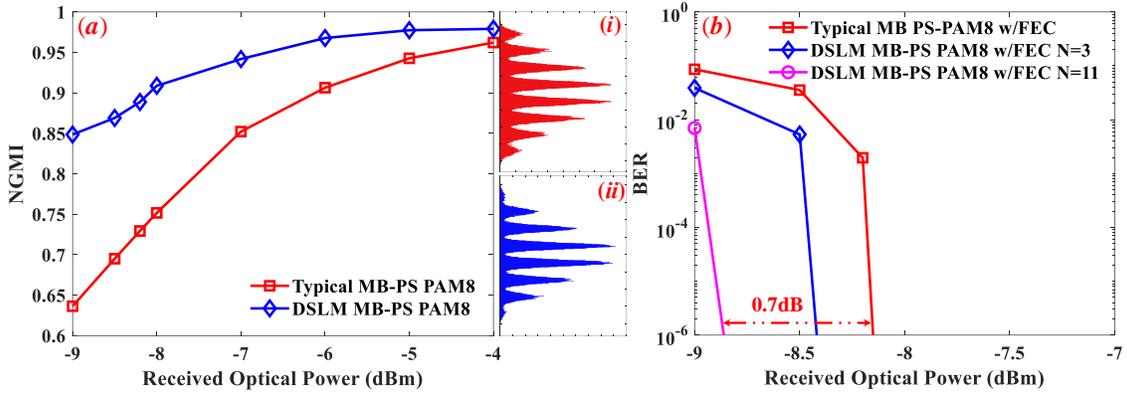

Fig. 20. (a) NGMI versus ROP of 56GBaud typical MB-PS PAM8 (H=2.8) and DSLM MB-PS PAM8. (b) BER versus ROP of typical MB-PS PAM8 and DSLM MB-PS PAM8 with turbo equalization. Symbol histograms of equalized (i) MB-PS PAM8 and (ii) DSLM MB-PS PAM8.

First of all, the proposed scheme is demonstrated in a 56GBaud uniform PAM8 system, which can be considered as the PAS module with a uniform output. Fig. 16(a) shows the distribution of the final transmitted PAM8 symbol when the memory length and the forbidden ratio of DSLM are 5 and 0.33, respectively. A non-uniform distribution is observed. More precisely, the entropy of the final transmitted PAM8 symbol is 2.97. The corresponding spectrum is depicted in Fig. 16(b). The eye diagrams of uniform PAM8 and DSLM PAM8 with a 100 up-sampling rate are depicted in Figs. 16(c) and 16(d). It can be found that the DSLM scheme effectively reduces fluctuation, although the entropy of the PAM8 symbol only decreases by 0.03 bits/symbol. Fig. 17 shows the normalized generalized mutual information (NGMI) versus the received optical power (ROP) of uniform and DSLM PAM8 signals. It is noteworthy that the NGMI for the DSLM scheme is computed between the received PAM8 symbols and the output PAM8 symbols of the DSLM unit, rather than the PAS unit. As shown in Fig. 17, the DSLM scheme effectively enhances the reliability of the system. About 1dB receiver sensitivity is observed. The eye diagrams of the recovered typical uniform PAM8 and DSLM PAM8 at −5dBm ROP are shown in the insets (i) and (ii) of Fig. 17, and a non-uniform distribution can be seen from the inset (ii). However, the NGMI calculated in the DSLM scheme does not take the bit ambiguity issue into consideration. We further evaluate the BER performance of the two schemes with turbo equalization. Fig. 18 shows BER versus the iterative number of turbo equalization under different ROP. At the ROP of −7dBm, the initial error is too large to achieve convergence. At −6.5dBm and −4dBm ROP, the required iterative number for error-free realization is 11 and 2, respectively. Fig. 19 shows the BER versus the ROP of the 56GBaud PAM8 system with and without DSLM. For a typical PAM8 system, the required ROP for error-free realization is −5.5dBm. For the DSLM PAM8 with turbo equalization (N=11), the required ROP for error-free transmission is −6.5dBm, achieving approximately 1dB receiver sensitivity improvement compared to the typical uniform PAM8 system.

Moreover, the proposed scheme is also verified in a 56GBaud MB-PS PAM8 system (H=2.8). As the input of DSLM is the MB-PS PAM8 symbol, the forbidden ratio should be large enough to obtain the shaping gain. In this case, the memory length and the forbidden ratio are 5 and 0.72. Fig. 20(a) depicts the measured NGMI versus the ROP of typical MB-PS PAM8 and DSLM MB-PS PAM8. Obviously, the envelope-controlled MB-PS PAM8 exhibits a great performance improvement compared to the typical MB-PS PAM8. The insets (i) and (ii) of Fig. 20(a) show the histograms of the equalized PAM8 symbol at −5dBm ROP. The distribution of DSLM MB-PS PAM8 exhibits a deviation from the traditional MB-PS PAM8. Finally, the BER versus the ROP of MB-PS PAM8 and DSLM MB-PS PAM8 after FEC decoding is depicted in Fig. 20(b). With the assistance of turbo equalization (11 iterations), the DSLM MB-PS PAM8 shows a 0.7dB receiver sensitivity improvement compared to the typical MB-PS PAM8.

## IV. Conclusion

In long-haul coherent systems, MB distribution coupled with PAS enables not only a 1.53dB maximum available shaping gain, but also a low complexity rate adaptability with fine granularity. However, due to the difference in the system model, the MB distribution is no longer the capacity-achieving distribution for IM-DD systems, resulting in an open question in this field. In this paper, we first numerically investigate the shaping gains of several commonly utilized distributions, including MB, IvMB, and BA-solved capacity-achieving distributions, for a memoryless PPC IM-DD system. According to numerical results, only a slight shaping gain is observed for the BA-solved distributions in the low AIR regions, and both the MB and IvMB distributions exhibit worse behavior than the BA-solved capacity-achieving distribution. Furthermore, a novel indirect PS technique is tailored for the PPC IM-DD system with memory effects, which aims to intentionally control the envelope of the transmitted signal, resulting in a higher robustness to the impairment induced by the system memory, such as overshoot, nonlinearity, and PAPR enhancement. The proposed scheme incorporates a DSLM algorithm at the transmitter, enabling an untypical bit-to-symbol mapping rule in which the current symbol is not only determined by the current bits pattern but also by previously generated symbols within a specified memory length. At the receiver side, a turbo equalization scheme with a modified M-BCJR algorithm is proposed to achieve the recovery of ambiguous bits induced by DSLM. The effectiveness of the proposed scheme is experimentally validated in a 56GBaud PAM8 system. The results indicate that about 1dB receiver sensitivity enhancement is achieved over a 2km SMF transmission. In addition, by integrating PAS with the proposed DSLM, both shaping gain and rate adaptability are achieved, enabling enhanced performance and flexibility in system design.